\documentclass[3p,times,authoryear]{elsarticle}
\usepackage{amsmath}
\usepackage{amsthm}
\usepackage{comment}
\usepackage{url}
\usepackage{graphicx} 
\usepackage[utf8]{inputenc}
\usepackage[capitalise]{cleveref}
\usepackage{xcolor}
\usepackage{lscape}
\usepackage{color}
\usepackage{soul}
\usepackage[figuresright]{rotating}
\usepackage{appendix}
\usepackage{subcaption}

\begin{document}

\begin{frontmatter}

\title{Understanding Changes in Travel Patterns during the COVID-19 Outbreak in the Three Major Metropolitan Areas of Japan}

\address[tokyoUni]{Department of Civil and Environmental Engineering, School of Environment and Society, Tokyo Institute of Technology, \\ M1-11, 2-12-1, O-Okayama, Meguro, Tokyo, 152-8552, Japan}
\address[ut]{Department of Civil Engineering, School of Engineering, The University of Tokyo}
\address[ub]{Department of Mathematics, University at Buffalo}

\author[tokyoUni,ut]{Takao Dantsuji\corref{cor}}\ead{t.dantsuji@plan.cv.titech.ac.jp}\cortext[cor]{Corresponding author.}
\author[ub,tokyoUni]{Kashin Sugishita}\ead{kashinsu@buffalo.edu}
\author[ut,tokyoUni]{Daisuke Fukuda}\ead{fukuda@civil.t.u-tokyo.ac.jp}

\begin{abstract}\quad 
Unlike the lockdown measures taken in some countries or cities, the Japanese  government declared a ``State of Emergency" (SOE) under which people were only requested to reduce their contact with other people by at least 70 \%, while some local governments also implemented their own mobility-reduction measures that had no legal basis. The effects of these measures are still unclear. Thus, in this study, we investigate changes in travel patterns in response to the COVID-19 outbreak and related policy measures in Japan using longitudinal aggregated mobile phone data. Specifically, we consider daily travel patterns as networks and analyze their structural changes by applying a framework for analyzing temporal networks used in network science. The cluster analysis with the network similarity measures across different dates showed that there are six main types of mobility patterns in the three major metropolitan areas of Japan: (I) weekends and holidays prior to the COVID-19 outbreak, (II) weekdays prior to the COVID-19 outbreak, (III) weekends and holidays before and after the SOE, (IV) weekdays before and after the SOE, (V) weekends and holidays during the SOE, and (VI) weekdays during the SOE. It was also found that travel patterns might have started to change from March 2020, when most schools were closed, and that the mobility patterns after the SOE returned to those prior to the SOE. Interestingly, we found that after the lifting of the SOE, travel patterns remained similar to those during the SOE for a few days, suggesting the possibility that self-restraint continued after the lifting of the SOE. Moreover, in the case of the Nagoya metropolitan area, we found that people voluntarily changed their travel patterns when the number of cases increased. 
\end{abstract} 


\end{frontmatter}

\section{Introduction}
A cluster of cases of what was initially believed to be pneumonia was reported in Wuhan on 31 December 2019, but this was later identified as a new coronavirus called COVID-19. As of 26 October 2020, COVID-19 has become a global pandemic, with the numbers of cases and deaths reaching more than 43 million and one million, respectively. In an attempt to mitigate the spread of COVID-19, numerous countries implemented a variety of mobility restriction measures, resulting in significant changes to people’s lifestyles. Many people commenced working from home, and refrained from traveling for non-essential activities. China was the first country to implement travel restrictions, and it has been reported that the number of cases in cities that implemented travel restrictions earlier was fewer than that in cities that did so later \citep{tian2020investigation}. Travel restrictions are clearly an important means of controlling the spread of COVID-19. Following China’s lead, many countries and cities such as France and New York ordered people to avoid unessential travel during their lockdowns. However, while the spread of COVID-19 was controlled during the lockdowns, the number of cases started to rise again after the lockdown measures were eased. While lockdowns are one of the most effective means of controlling the spread of COVID-19, the economic damage has been immense. \cite{baek2020unemployment} claim that the lockdowns in the United States caused significant job losses.  Thus, as a less strict measure, on 7 April 2020 the Japanese government declared a ``State of Emergency'' (SOE) for seven major prefectures, and then extended it to all prefectures on 16 April. However, because people were only ``requested" to reduce their interactions with other people by at least 70 \%, there were no penalties or punishments if they ignored the request. In addition to the SOE, local governments implemented their own measures against COVID-19 (see Figure \ref{fig:time}). However, the effects of the SOE and other measures on changes in travel patterns remain unclear. To identify the appropriate balance between mitigating the spread of COVID-19 and economic damage, the changes in travel behavior caused by less strict measures such as those applied in Japan should be investigated to assist policy-makers in the future.

Recent developments in information and communications technology have enabled the collection of more and more data in relation to human mobility, and mobile phone data provide a tool for understanding travel patterns. Recently, aggregated mobile phone data have been used for urban-scale travel pattern analyses such as traffic demand estimation \citep{iqbal2014development, alexander2015origin, janzen2018closer, ge2016updating,bachir2019inferring} and traffic or travel behavior model calibration \citep{sawada2019update,paipuri2020estimating}. A comprehensive literature review of the application of mobile phone data to travel behavior analysis can be found in \cite{wang2018applying}. Mobile phone data have also been used for analyses of travel behavior during the COVID-19 pandemic. For instance, \cite{arimura2020changes} investigated population changes in the central business district of Sapporo city in Japan using mobile phone data and found that population densities fell by up to 90 \% in some previously busy areas.  \cite{santamaria2020measuring} investigated several mobility indices such as internal and inward demand based on the origin--destination matrix obtained using mobile phone data. \cite{heiler2020country} analyzed the influence of the lockdown in Austria using mobile phone data and various measures such as a clustering index. However, no one has investigated how travel patterns changed over time and whether they returned to those prior to the implementation of measures such as lockdowns.

Thus, in this study, we aim to identify changes in travel patterns during the COVID-19 outbreak in Japan and discuss the effects of the non-compulsory measures that were imposed. To achieve these objectives, we consider daily travel patterns as temporal networks, wherein the network structure changes over time \citep{holme2012temporal, holme2015modern, holme2019temporal, masuda2016guidance}, and apply the framework proposed by \cite{masuda2019detecting}. Using this framework, we can analyze structural changes in temporal networks from the viewpoint of recurrence, which implies that the network structure returns to one relatively close to that in the past. Recurrence has been studied in the context of various temporal networks such as the human brain \citep{lopes2020recurrence}, hashtag use on Twitter \citep{cruickshank2020characterizing}, and air transport \citep{sugishita2020recurrence}. From the viewpoint of recurrence, we study how travel patterns changed during the COVID-19 outbreak in the three major metropolitan areas of Japan, and conjecture that the SOE dramatically changed people’s travel patterns. However, it is possible that people may have voluntarily changed their travel patterns even before the SOE was declared. Further, travel patterns following the lifting of the SOE may have returned to those prior to the COVID-19 outbreak. 

The rest of the paper is organized as follows. In Section 2, we describe the dataset we used, including the study areas, events and measures, and the type of mobile phone data obtained. In Section 3, we present the method used to analyze travel patterns. In Section 4, we present the results and discuss their implications. In Section 5, we present our conclusions and discuss possible directions for future research.

\section{Dataset}
\subsection{Study Areas}
\begin{sidewaysfigure}
\centering
\includegraphics[width=\columnwidth]{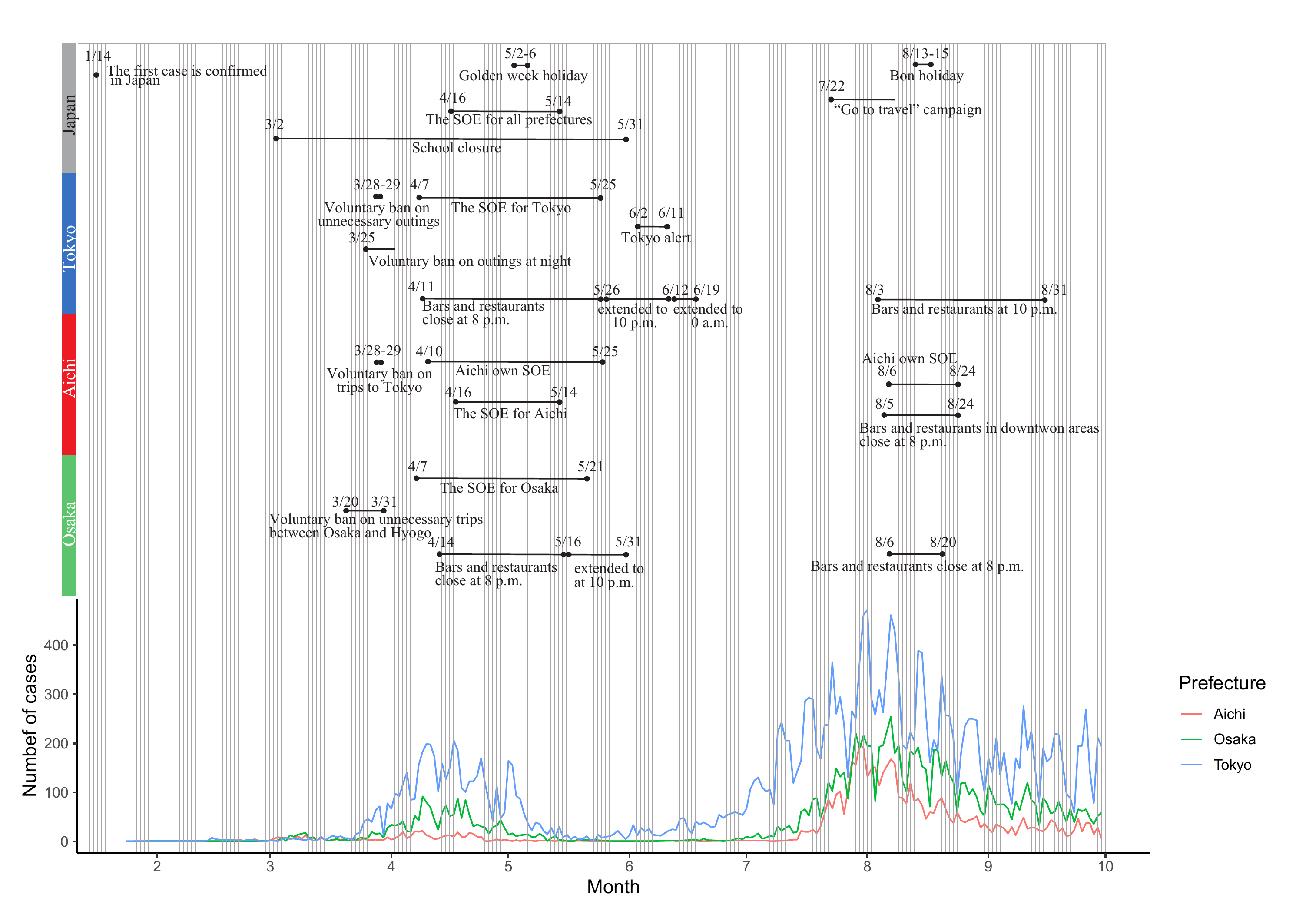}
\caption{Time series of events related to the COVID-19 pandemic showing the numbers of cases in Tokyo, Aichi, and Osaka, which are the core prefectures of the T-UEA, N-UEA, and O-UEA, respectively. }
\label{fig:time}
\end{sidewaysfigure}

The study areas are the Tokyo, Nagoya, and Osaka urban employment areas (UEAs) as defined by \cite{kanemoto}, which are the three largest metropolitan areas in Japan. Regarding the definition of a UEA, a municipality belongs to a UEA if more than 10 \% of the workers who live in that municipality commute to the core of the UEA, which is the densely inhabited district where the population is greater than 50,000. The areas of the Tokyo UEA (T-UEA), Nagoya UEA (N-UEA), and Osaka UEA (O-UEA) are approximately 9,510, 3,840, and 3,760  [${\rm km}^2$], respectively, and the populations of the T-UEA, N-UEA, and O-UEA  are approximately 35, 12, and 6 million, respectively. We chose these three UEAs as the study areas because the main prefectures (i.e., Tokyo for the T-UEA, Aichi for the N-UEA, and Osaka for the O-UEA) implemented different measures in response to the COVID-19 outbreak, as shown in Figure \ref{fig:time}. After the first case of COVID-19 in Japan was confirmed on 14 January 2020, the Japanese government ordered all schools to be closed on 2 March as the first step toward controlling the spread of COVID-19. Then, on 7 April, the government declared a SOE for Tokyo, Kanagawa, Chiba (in the T-UEA), Osaka, Hyogo (in the O-UEA), and Fukuoka. When it became clear that the number of cases was not falling, the SOE was extended to all prefectures on 16 April. Then, following the first peak, the SOE was lifted on 14 May for all prefectures with the exception of some such as Tokyo and Osaka, where the number of cases was relatively high. The SOE was finally lifted in all prefectures on 25 May, and schools were allowed to reopen. The Ministry of Education, Culture, Sports, Science and Technology reported that 93 \% of public schools were closed on 22 April, 87 \% on 11 May, and only 1 \% on 1 June \citep{MEXT2020}. On 22 July, the Japanese government commenced the ``Go To Travel" campaign, whereby it offered discounts for domestic travel in an attempt to boost the recovery of the Japanese economy. Meanwhile, the national government implemented a variety of measures, while local governments also tried to control the spread of COVID-19. When the number of cases increased rapidly at the end of March, the Tokyo local government requested that people avoid unnecessary outings at night and on weekends, while the Aichi local government requested that people avoid trips to Tokyo. The Osaka local government also requested that people avoid trips between Osaka and Hyogo prefectures between 20 and 31 March because the number of cases was increasing in both prefectures. Later, even though Aichi prefecture is one of the largest metropolitan areas in Japan, the national government did not declare an SOE for Aichi at the same time as it did for Tokyo and Osaka prefectures. Thus, the Aichi local government declared its own SOE on 10 April even though it had no legal basis, and declared another SOE on 6 August when the number of cases rapidly increased. Because the number of COVID-19 cases started to increase again at the end of May, the Tokyo local government declared ``Tokyo Alert" on 2 June, which is an Tokyo own warning system that works when some criteria such that the daily average of the number of cases in the past seven days is more than 20 are satisfied and it was lifted on 11 June. Besides those measures, there are two long holidays in Japan, which are Golden week holiday from 2 May to 6 May and Bon holiday from 13 August to 15 August. These long holidays could also changes the travel patterns. 

\begin{figure}[h]
\centering
\includegraphics[width=\columnwidth]{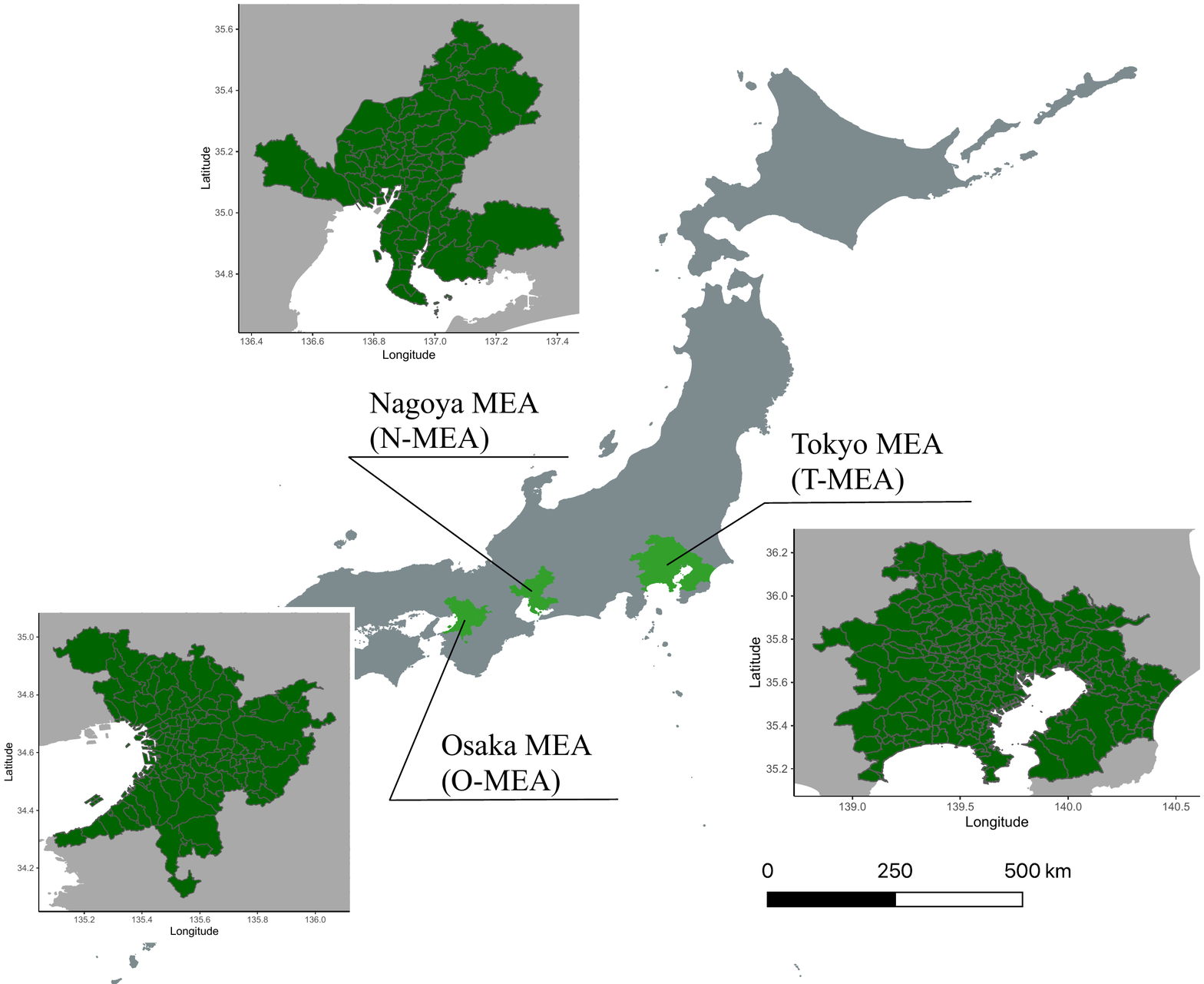}
\caption{Locations of the T-UEA, N-UEA, and O-UEA.}
\label{fig:UEA_location}
\end{figure}

\subsection{Mobile Spatial Statistics}
In this study, we used aggregated mobile phone data termed ``Mobile Spatial Statistics" (MSS) from 1 January to 30 September of 2020\footnote{Note that the data on 19 February is removed due to the system issue.} provided by Docomo Insight Marketing, Inc. NTT DOCOMO, Inc. is one of Japan’s largest mobile phone providers, and its MSS dataset contains hourly data on the age, sex, and city of residence of the population in all areas of Japan (divided into a 500 m square grid). Because there are several mobile phone providers, the populations in various areas are estimated based on the market share of various mobile phone providers. The hourly population in a given area is estimated using mobile phone data as follows. The number of people in an area is estimated by considering how long users stay in the area. For instance, if a user stays in an area for 15 minutes, they are regarded as representing 0.25 of a person in that area. Then, after the number of people in the area is aggregated, the hourly population is obtained by scaling using the market share of people of their age and sex who live in that city and use NTT DOCOMO, Inc. as their mobile phone provider. This information was obtained from the customer information held by NTT DOCOMO, Inc. and the Basic Resident Registration System of Japan’s Ministry of Internal Affairs and Communications. Because MSS data include information on the users' age, sex, and city of residence, they can be used for various travel pattern analyses including updating the parameters of the travel behavior model \citep{sawada2019update}. Note that we use data for residential cities that are in the same UEAs, as we are focused on changes in travel patterns related to daily activities such as work and shopping at urban scale. Thus, travel for tourism or business trips outside the UEAs are beyond the scope of this study. However, if we were to expand the study area to the entire country, it would be easy to include such travel. 

\section{Methodology}
\subsection{Mobility pattern networks}
\begin{figure}[t]
\centering
\includegraphics[width=\columnwidth]{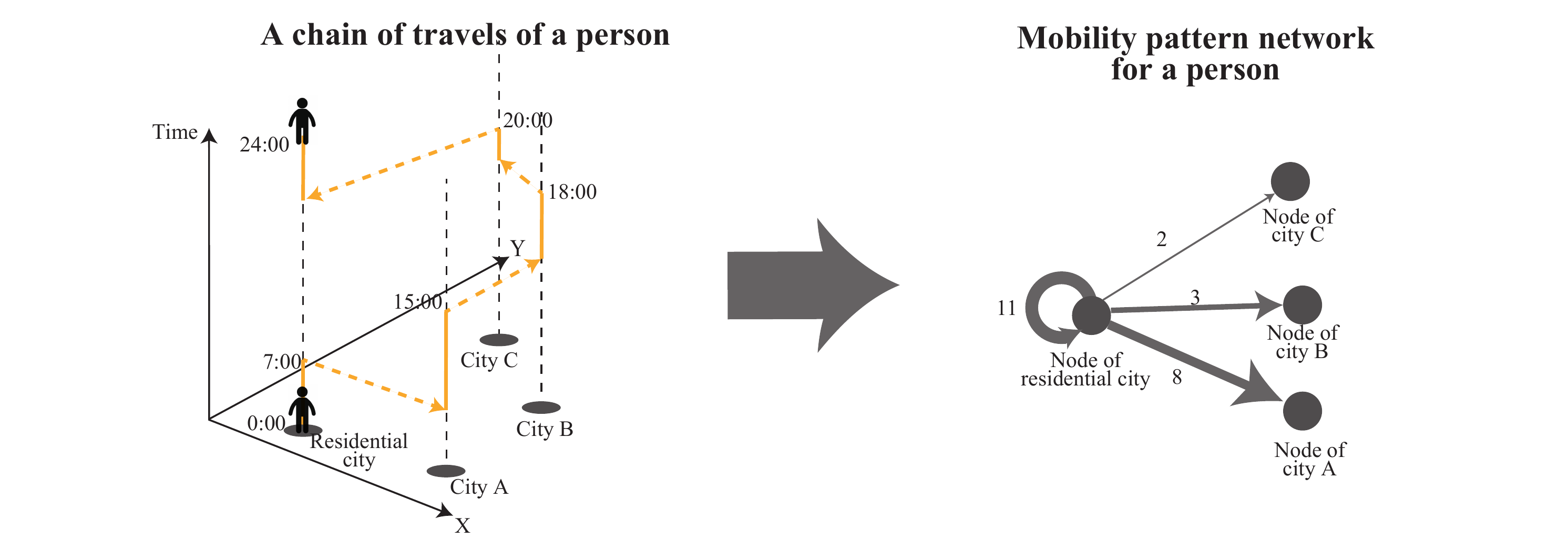}
\caption{Graphical representation of the concept of a mobility pattern network. }
\label{fig:gmpn}
\end{figure}

Because MSS represent aggregated mobile phone data, it is not possible to obtain the detailed trajectories of the various travel behaviors. Therefore, to understand the changes in travel patterns in response to the COVID-19 pandemic, we constructed networks (referred to as mobility pattern networks) from the MSS that include information regarding the city of residence of the users. To construct the mobility pattern networks, we aggregated the data contained in the 500 square m grid at the city level, as the information regarding the place of residence is provided at the city level. We treated the various residential cities as nodes from which the users start their trips on a given day. Because the cities are regarded as nodes, the aggregated population movement from one residential city to another city on a given day is considered to represent the weight of the link between those two cities. A graphical representation of the concept of the mobility pattern network of a person is shown in Figure \ref{fig:gmpn}. For instance, a person travels to city A, where her/his workplace is located, at 7 am, and then makes a business trip to city B at 3 pm. After work, the person goes to a gym in city C at 6 pm, and then returns home at 8 pm. Since the time this person spends in city A, city B, and city C is eight, three, and two hours, respectively, the weights of the links from the residential city node to those of city A, city B, and city C are 8, 3, and 2, respectively. Moreover, as the time spent in the residential city is 11 hours, the weight of a link from and to the residential city node is 11. 

As we construct the daily mobility pattern networks for the UEAs, the weight of link $(i,j)$ on day $t_1$ is given by $D_{i,j}^{t_1} = \sum_{h \in H} D_{i,j}^{h, t_1}\,,$
where $D_{i,j}^{h, t_1}$ is the number of people in city $j$ who live in city $i$ at time of day $h$ on day $t_1$, and $H$ is the set of times of day (i.e., \{1, 2, ..., 24\}). Thus, the total weight of all the links in a mobility pattern network can be estimated by multiplying the total estimated population by 24 hours. Thus, it can be regarded as a daily record of people’s activities, including where they spend their time and how long they spend in each location. 
Although travel patterns are traditionally presented as origin--destination (OD) matrices  \citep{ge2016updating}, we constructed mobility pattern networks instead of using OD estimations because a mobility pattern network can capture  the duration of people's various activities and their trip chains. These aspects are important in relation to understanding travel behavior in circumstances such as the COVID-19 pandemic. For instance, some office workers might work from home in the morning, and then travel to their office in the middle of the day to avoid the overcrowded trains during the morning peak period. This seldom occurs in normal circumstances, especially in Japan. Detailed properties of people’s travel behavior such as the purpose of their travel cannot be obtained from mobility pattern networks. However, changes in travel behavior are reflected by changes in mobility pattern networks.

\subsection{Recurrence in mobility pattern networks}
We analyzed structural changes in the mobility pattern networks using the framework for analyzing temporal networks \citep{masuda2019detecting} with a distance measure for weighted networks. Let $G_{t_1}(V, E)$ be the weighted and directed network of the mobility pattern on day $t_1$, where $V$ is the set of nodes and $E$ is the set of weighted and directed links. To obtain the dissimilarities among the mobility pattern networks, we use the similarity function for the links between the nodes. The similarity function of link $(i,j)$ between day $t_1$ and day $t_2$ is defined as
\begin{eqnarray}
w_{ij} (t_1, t_2) = \exp\left\{-\left(\frac{D_{ij}^{t_1}-D_{ij}^{t_2}}{\theta}\right)^2\right\} \,, \, \, \forall \ ij \in  E(G_{t_1} \cup G_{t_2})\,,
\end{eqnarray}
where $D_{ij}^{t_1}$ is the weight of link $(i,j)$, which represents the number of people who travel to the destination of link $(i,j)$, $\theta$ is the scale (bandwidth) parameter that adjusts the criteria of the outliers, and $E(G_{t_1} \cup G_{t_2})$ is a set of links that exist in either network $G_{t_1}$ or network $G_{t_2}$. Note that the weight of link $(i,j)$ on day $t$ is set to zero when there is no link on day $t$.  Because the weights of the links are important, we use the weighted similarity function. This is a Gaussian-type probability density function that is also used for similarity functions to partition transportation road networks into homogenous sub-networks \citep{ji2012spatial,dantsuji2019cross}. The values of the similarity function are between 0 and 1, with higher values representing greater similarity between networks.
To compare two mobility pattern networks, $G_{t_1}$ and $G_{t_2}$, we define the dissimilarity function  as follows:  
\begin{eqnarray}
d(G_{t_1},G_{t_2}) = 1 - \frac{\sum_{ij \in E(G_{t_1} \cup G_{t_2})} w_{ij} (t_1, t_2)}{|E(G_{t_1} \cup G_{t_2})|} \,.
\end{eqnarray}
To identify the mobility pattern changes over time, we apply the hierarchical clustering method to the distance matrix, whereby each element represents the value of the dissimilarity function between two given days \citep{masuda2019detecting}. Note that the distance used to merge the clusters in the algorithms of the hierarchical clustering method is estimated using Ward's method \citep{ward1963hierarchical}, and the number of optimal clusters is determined using Dunn's index \citep{dunn1973fuzzy}. 

\section{Results}
\subsection{Time series of the mobility pattern networks}
\begin{figure}[p]
	\begin{minipage}{\columnwidth}
	\end{minipage}
	\begin{minipage}{\columnwidth}
		\centering
		\includegraphics[width=\columnwidth]{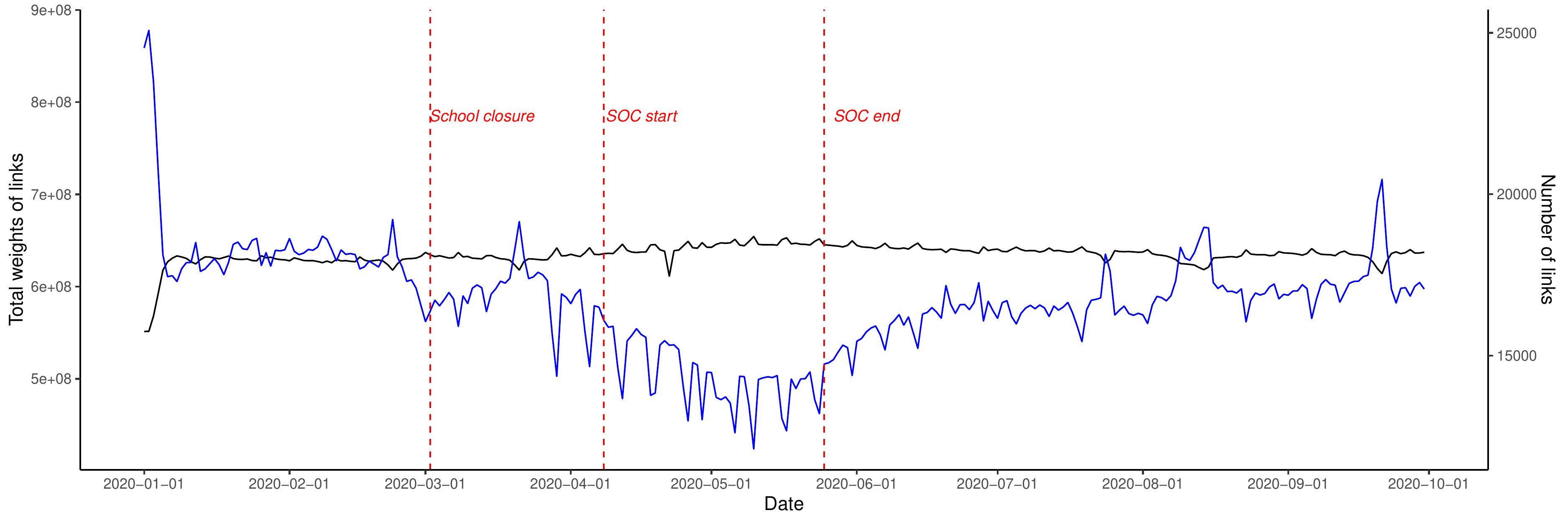}
		\subcaption{T-UEA}
	\end{minipage}
		\begin{minipage}{\columnwidth}
		\centering
		\includegraphics[width=\columnwidth]{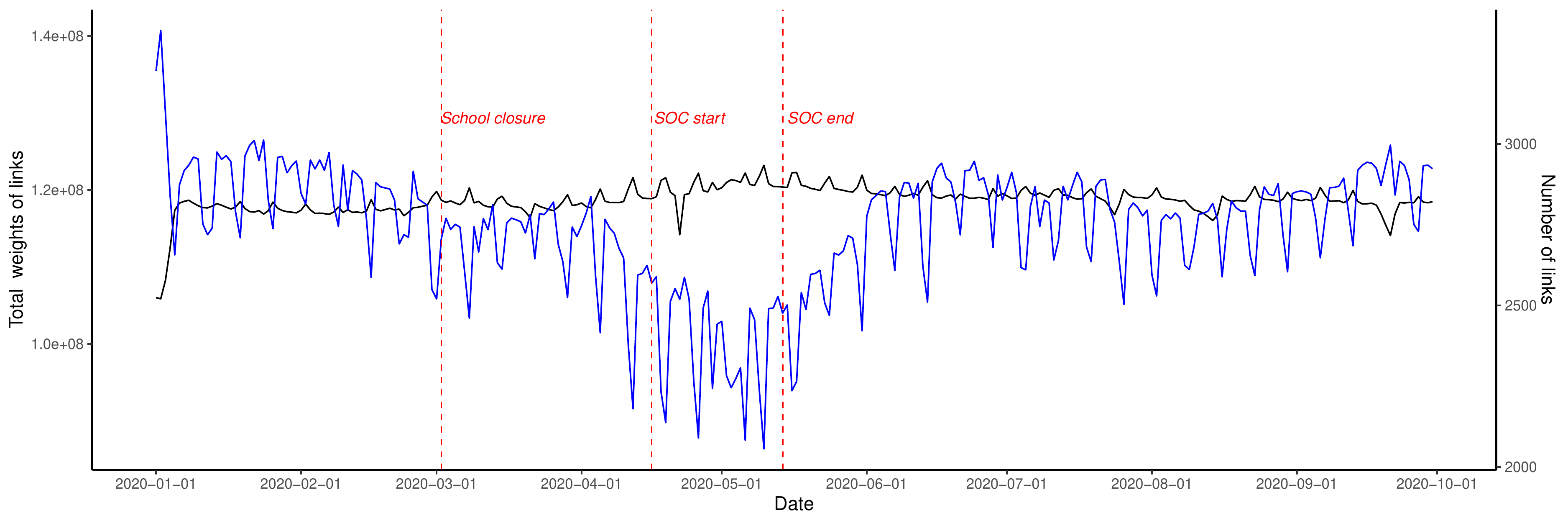}
		\subcaption{N-UEA}
	\end{minipage}
		\begin{minipage}{\columnwidth}
		\centering
		\includegraphics[width=\columnwidth]{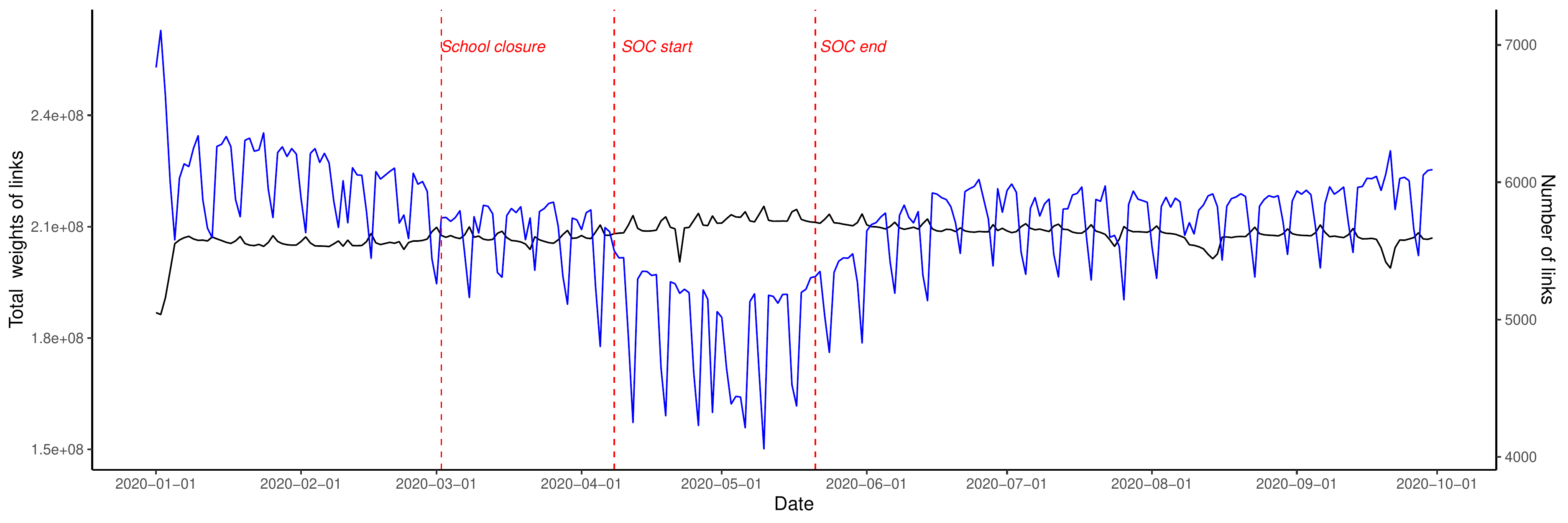}
		\subcaption{O-UEA}
	\end{minipage}
\caption{Time series for total weights and the number of links. Black and blue lines represent total weights and the number of links, respectively.}
	\label{fig:TS}
\end{figure}

 We first investigate how the number of links and total weights of the mobility pattern networks change over time, as shown in Figure \ref{fig:TS}. It can be seen that the total weights of the mobility pattern networks do not change significantly except for during the New Year holiday period. The reduction in the total weights during the New Year holiday period was caused by people returning home or traveling beyond these UEAs. Conversely, the number of links in the mobility pattern networks changes significantly over various days. After the government ordered schools to close in March, the number of links fell in all three UEAs. Interestingly, the number of links gradually decreased around the end of March prior to the declaration of the SOE. This could have been because on 26 March, the prefecture governments (Tokyo, Saitama, Chiba, and Yamanashi) in the T-UEA requested that people avoid any unnecessary outings and work from home, which changed their travel behavior. The number of links was much lower during the SOE than during other time periods in all three UEAs. This could have been because people voluntarily limited travel from their city of residence during the SOE. After the SOE was lifted, it took roughly a month for the number of links to return to the level that existed during the period between the closure of schools and the declaration of the SOE. This change led to an increase in the number of COVID-19 cases, as shown in Figure \ref{fig:TS}. The number of links in the N-UEA fell slightly around the beginning of August, but those in the T-UEA and O-UEA did not fall. This could have been the result of Aichi's self-imposed SOE, although the following analysis shows that the changes in travel patterns had already begun before Aichi declared its own SOE. 

\begin{figure}[p]
\begin{minipage}{0.49\columnwidth}
		\centering
		\includegraphics[width=\columnwidth]{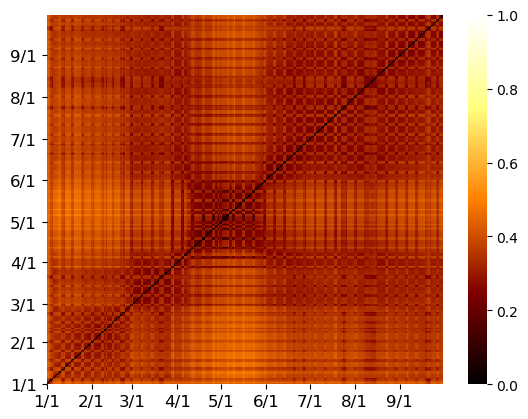}
		\subcaption{T-UEA}
	\end{minipage}
\begin{minipage}{0.49\columnwidth}
		\centering
		\includegraphics[width=\columnwidth]{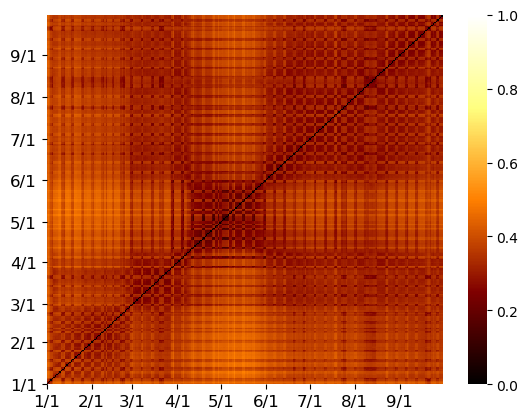}
		\subcaption{N-UEA}
	\end{minipage}
	\begin{minipage}{0.49\columnwidth}
		\centering
		\includegraphics[width=\columnwidth]{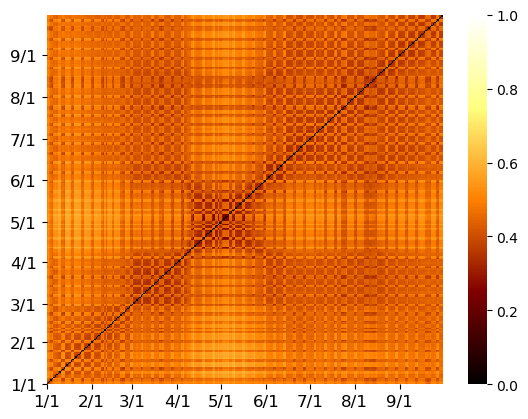}
		\subcaption{O-UEA}
	\end{minipage}
	\caption{Distance matrices for the T-UEA, N-UEA, and O-UEA.}
	\label{fig:dm}
\end{figure}

\subsection{Distance matrices}
Figure \ref{fig:dm} shows the distance matrices for the T-UEA, N-UEA, and O-UEA. Darker shading indicates more similar mobility pattern networks over two days, while lighter shading indicates less similar networks. Note that the values of the scale parameter $\theta$ are set so that the similarity of the median of the differences in link weights between days for all OD pairs in the UEAs is 0.5. Thus, $\theta$ = 55, 138, and 108 for the T-UEA, N-UEA, and O-UEA, respectively. From the distance matrices, it can be seen that there were clear grid patterns prior to the COVID-19 outbreak (i.e., from January to the beginning of March) in all of the UEAs. This indicates that the dissimilarity function was able to capture the differences in mobility pattern networks between weekdays and weekends. It can also be seen that the mobility pattern networks from April to June are quite different, as the values of the dissimilarity function during that period are higher than those during other periods. We expect that the SOE changed the mobility patterns significantly compared with those shown in the distance matrices. 

\subsection{Clustering results}
Based on the distance matrices shown above, hierarchical clustering was used to investigate the changes in the mobility pattern networks. Roughly speaking, we can classify the mobility pattern networks into six types:(I) weekends and holidays prior to the COVID-19 outbreak, (II) weekdays prior to the COVID-19 outbreak, (III) weekends and holidays before and after the SOE, (IV) weekdays before and after the SOE, (V) weekends and holidays during the SOE, and (VI) weekdays during the SOE. 
\subsubsection{T-UEA}
\begin{figure}[p]
		\centering
		\includegraphics[width=\columnwidth]{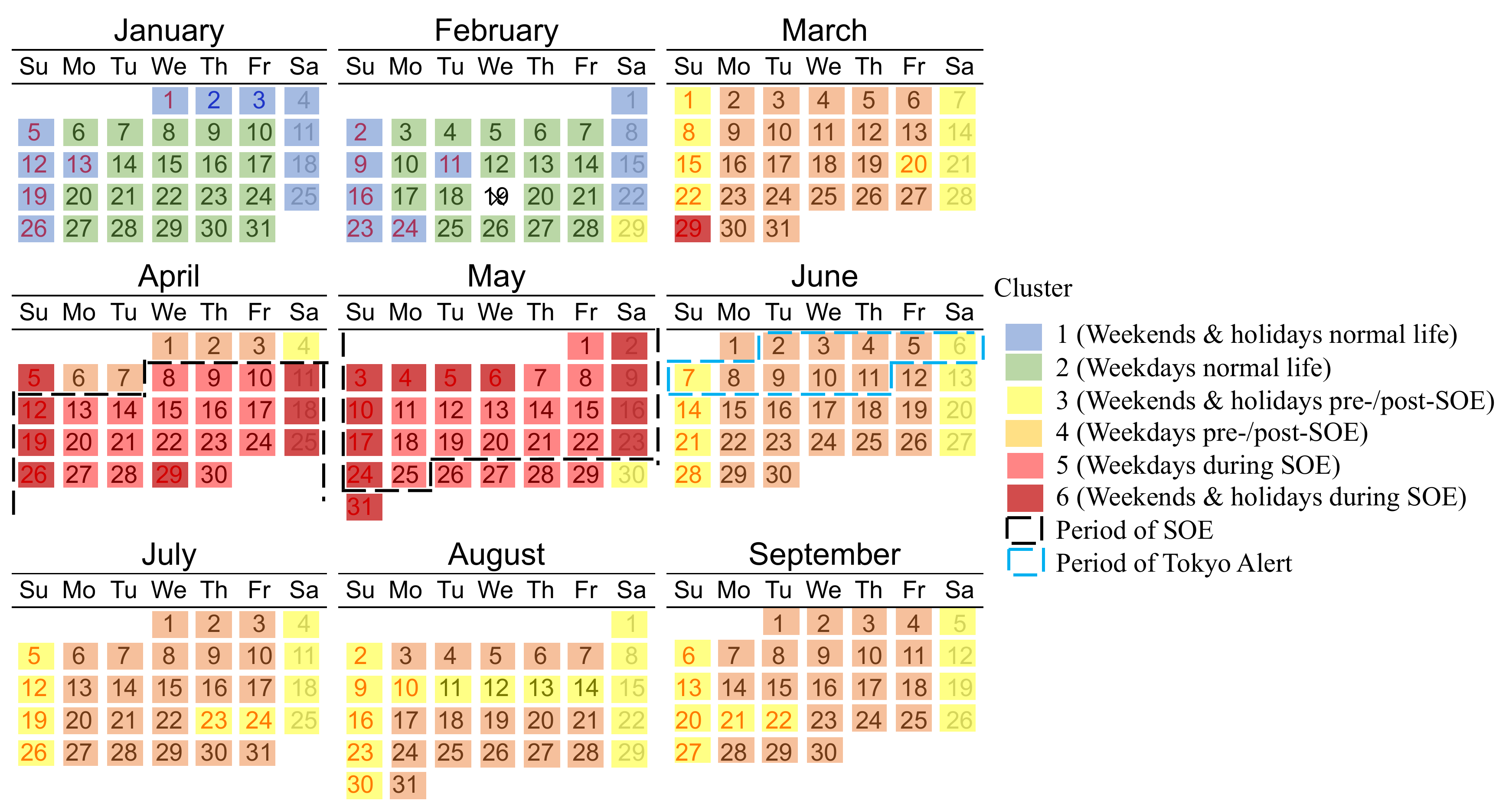}
		\caption{Clustering results for the T-UEA.} 
		\label{fig:cr_t}
		\vspace{2em}
				\centering
		\includegraphics[width=\columnwidth]{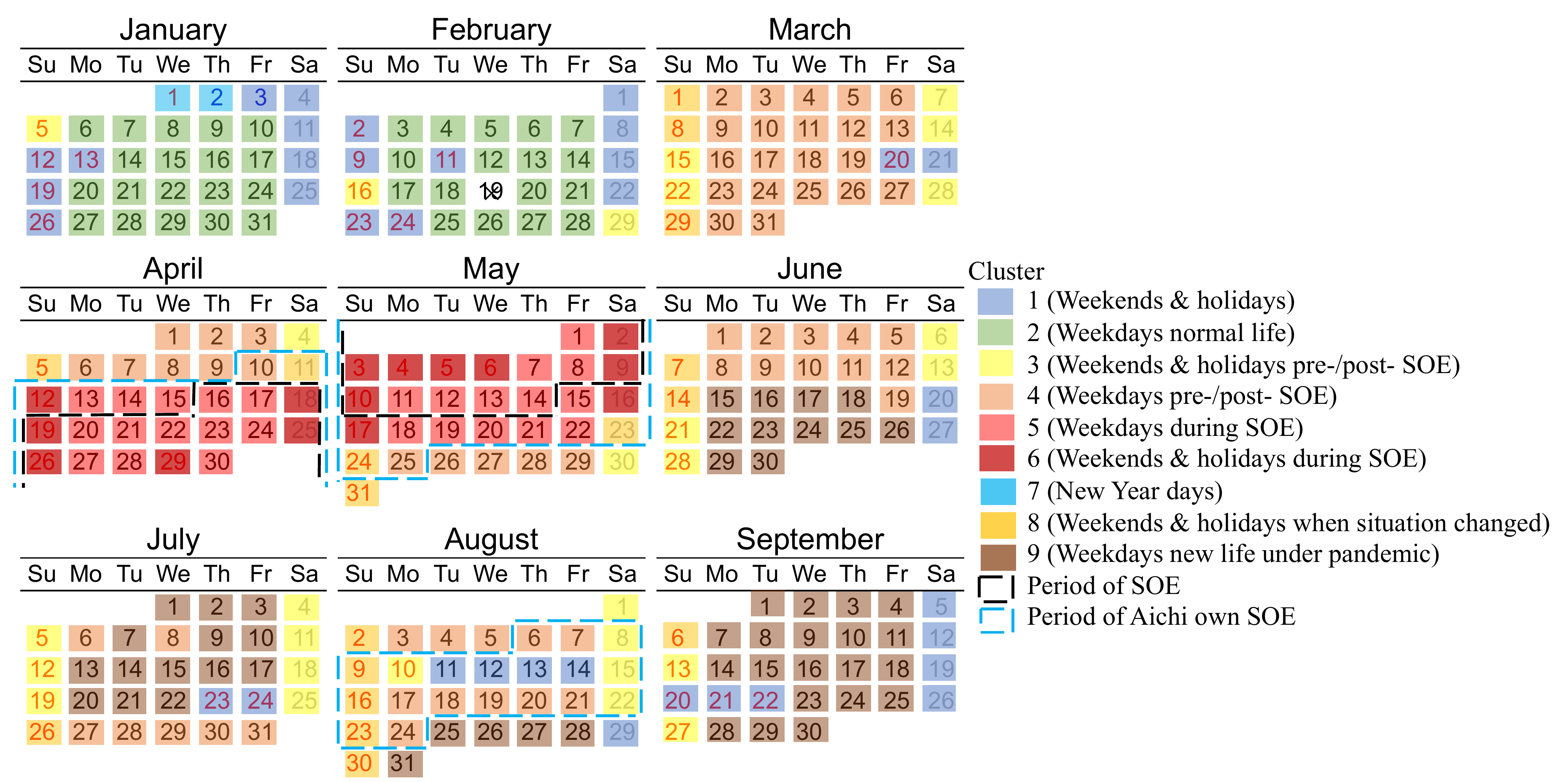}
		\caption{Clustering results for the N-UEA.} 
			\label{fig:cr_n}
\end{figure}

We obtained six clusters for the T-UEA, as shown in Figure \ref{fig:cr_t}. Cluster 1 contains all weekends and holidays prior to March, and cluster 2 contains all weekdays prior to March. As the school closure was ordered in March, clusters 1 and 2 are labeled ``holidays \& weekends normal life" and ``weekdays normal life," respectively. Following the closure of the schools, the mobility pattern networks changed. Clusters 3 and 4 contain holidays and weekends, and weekdays after March except for the period during the SOE, respectively. One interesting finding is that following the lifting of the SOE, mobility patterns returned to those prior to the SOE. Thus, we labeled clusters 3 and 4 ``Weekends \& holidays pre/post-SOE" and ``Weekdays life pre/post-SOE," respectively. As most of the days in clusters 5 and 6 are weekdays, and weekends and holidays, respectively, during the SOE (7 April to 25 May), clusters 5 and 6 are labeled ``weekdays during SOE" and ``Weekends \& holidays during SOE," respectively. Even though the SOE for Tokyo prefecture was lifted on 25 May, the mobility pattern networks from 26 May to 29 May are included in cluster 5 because they are similar to those during the SOE. This could be because people were still exercising caution. We can also indicate that the impact of ``Tokyo alert" on travel pattern changes is not evident in the clustering results.

\subsubsection{N-UEA}
Nine clusters were detected for the N-UEA. Cluster 2 contains all weekdays prior to March. Weekdays, and weekends and holidays during the SOE are contained in clusters 5 and 6, respectively. Thus, clusters 2, 5, and 6 were labeled ``Weekdays normal life," ``weekdays during SOE," and ``Weekends \& holidays during SOE," similar to the results for the T-UEA. Like the T-UEA, even after the SOE was lifted on 14 May in Aichi prefecture, the mobility pattern networks remained similar to those during the SOE until 22 May. Cluster 1 (labeled ``Weekends \& Holidays outside SOE") includes not only weekends prior to March but also some weekends before and after the SOE. One possible reason is because the number of COVID-19 cases was lower than those in Tokyo and Osaka, as shown in Figure \ref{fig:time}, people might not have changed their travel behavior significantly. Clusters 3 and 8 (labeled ``Weekends \& holidays pre/post-SOE" and ``weekends \& holidays when situation changed") also contain weekends immediately before and after the SOE. The difference between these two clusters is that the days in cluster 8 tend to appear when the situation has changed (i.e., from normal life to life before the SOE).  People might also have waited to see what others were doing on days when the situation changed. The days in cluster 4 tend to appear just before and after the SOE, while cluster 9 includes days distant from the SOE. Thus, clusters 4 and 9 were labeled ``Weekdays pre/post-SOE" and ``Weekdays new life under pandemic," respectively. This is because it can be seen that people were more worried about the outbreak and refrained from outings near the period of the SOE, while cluster 9 can be seen as representing days on which people felt more able to undertake outings. In fact, when the number of cases increased to the same level as Osaka prefecture at the end of July (see Figure \ref{fig:time}), the mobility pattern networks returned to those of cluster 4, and then returned to those of cluster 9 on 25 August after the number of cases started to fall and Aichi’s SOE were removed. It can also be seen that people changed their mobility patterns prior to Aichi declaring an own SOE because the transition from cluster 9 to cluster 4 occurred on 27 July. 

\begin{figure}
		\centering
		\includegraphics[width=\columnwidth]{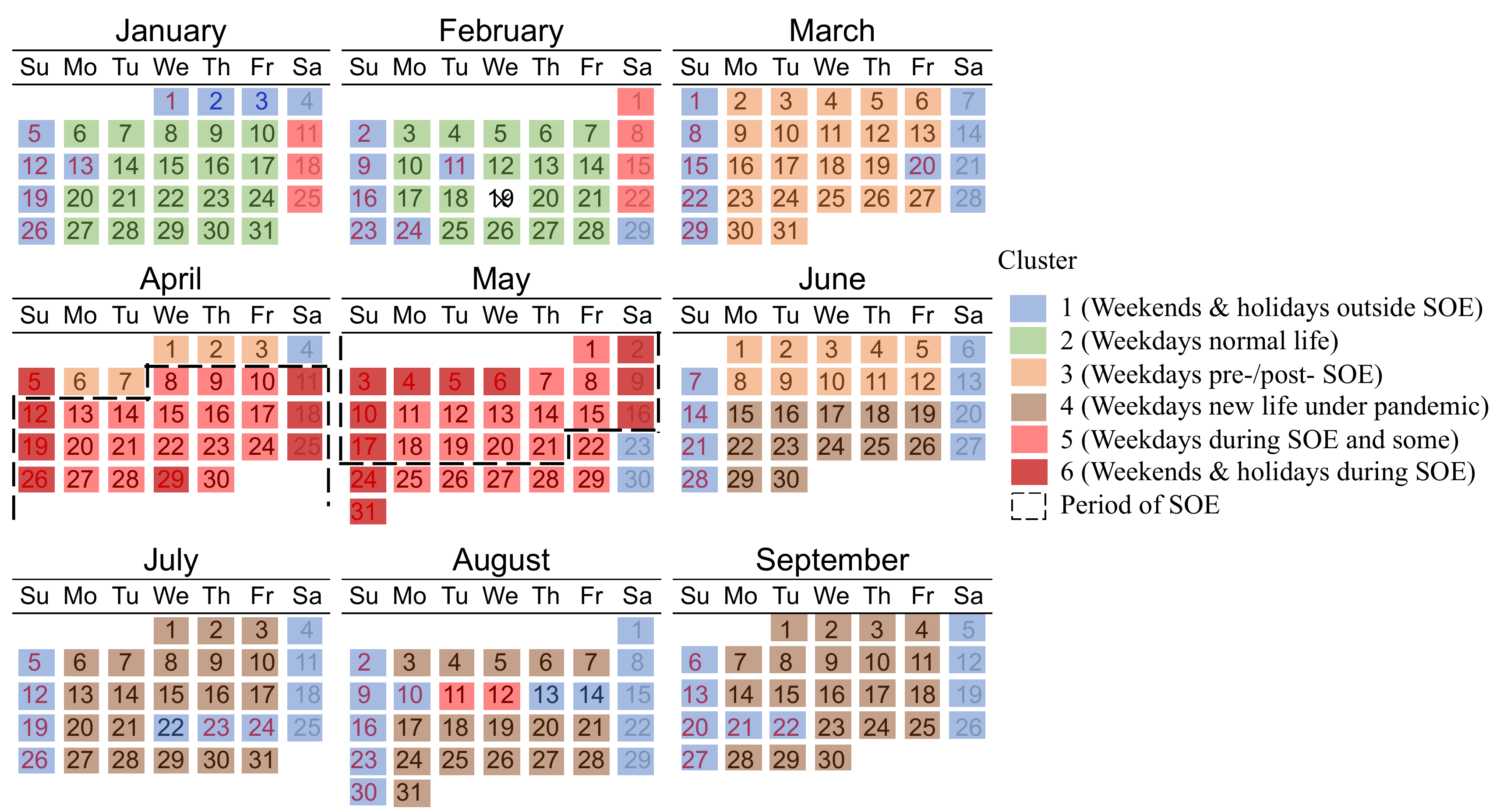}
        \caption{Clustering results for the O-UEA.}    
        		\label{fig:cr_o}
\end{figure}

\subsubsection{O-UEA}
The number of clusters for the O-UEA is the same as that for the T-MEA.  However, the characteristics of the clusters differ. For instance, most weekdays and holidays outside the SOE period are contained in cluster 1 (labeled ``weekends and holidays outside SOE"). Moreover, some weekdays and holidays are included in the same cluster as weekdays during the SOE (i.e., cluster 5 is labeled ``weekdays during SOE and some"). Similar to the T-UEA and the N-UEA, the mobility pattern networks for weekdays changed in March. In addition, even after the SOE was lifted on 21 May, the same mobility pattern remained in cluster 5 (``weekdays during SOE"). Then, from 15 June, the mobility pattern network changed to that of cluster 4 (``weekdays new life under pandemic"), which can be interpreted in the same way as cluster 9 for the N-UEA, that is, people felt more able to undertake outings. Although the mobility pattern network in the case of the N-UEA returned to that of cluster 4, wherein people were more worried about COVID-19, this did not happen in the case of the O-UEA.  

\subsubsection{Dendrograms}
\begin{figure}[p]
	\centering
	\begin{minipage}{\columnwidth}
		\centering
		\includegraphics[width=\columnwidth]{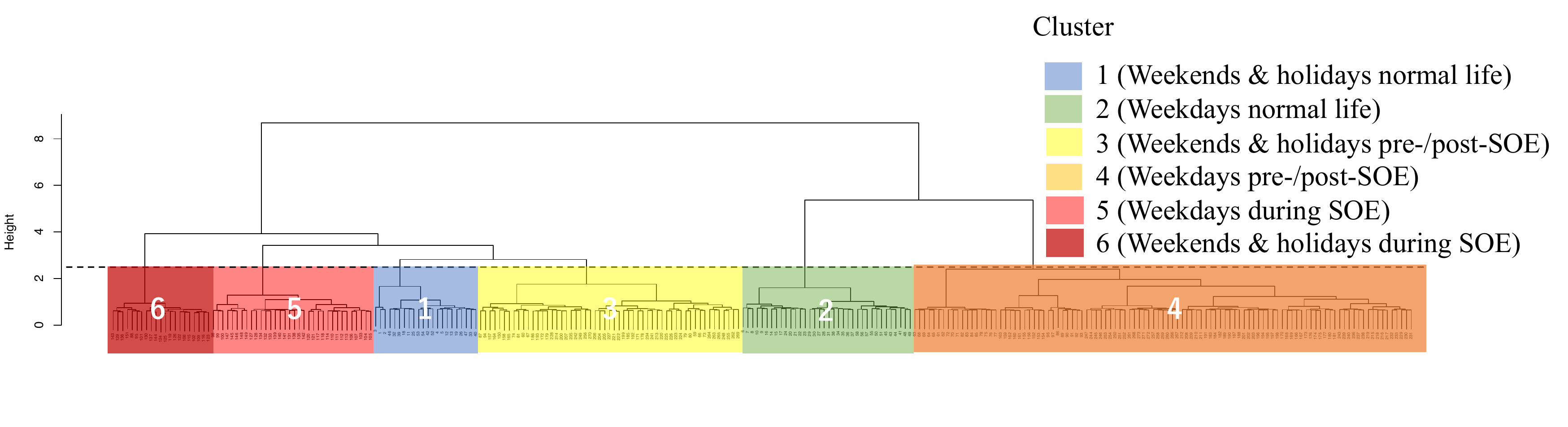}
		\subcaption{T-UEA}
		\label{fig:comparison}
	\end{minipage}
	\begin{minipage}{\columnwidth}
		\centering
		\includegraphics[width=\columnwidth]{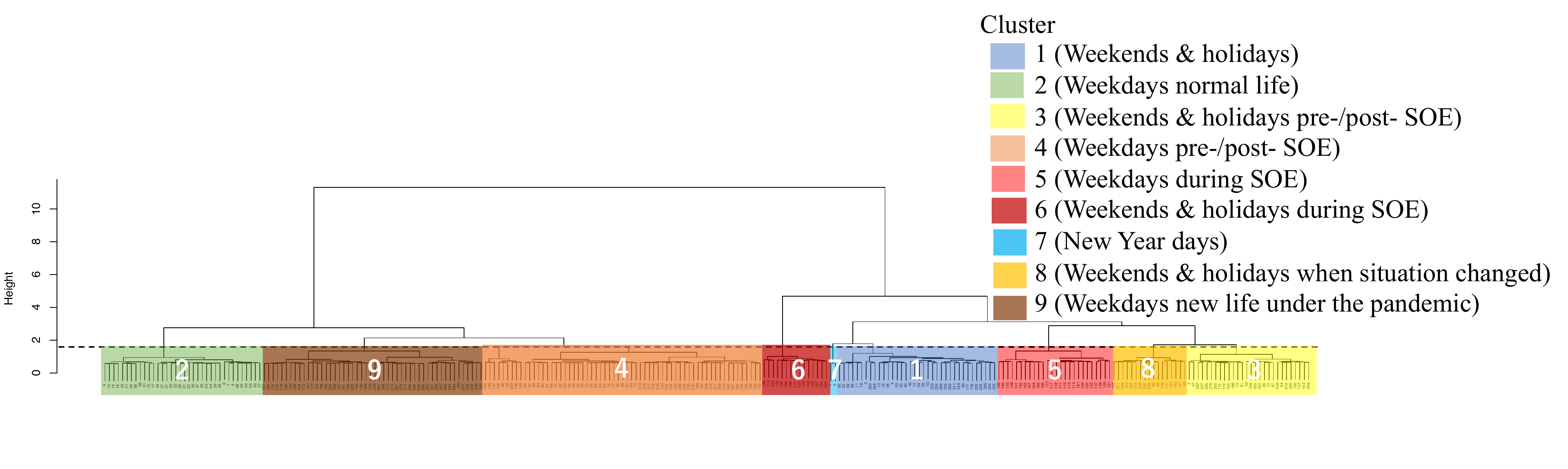}
		\subcaption{N-UEA}
	\end{minipage}
		\begin{minipage}{\columnwidth}
		\centering
		\includegraphics[width=\columnwidth]{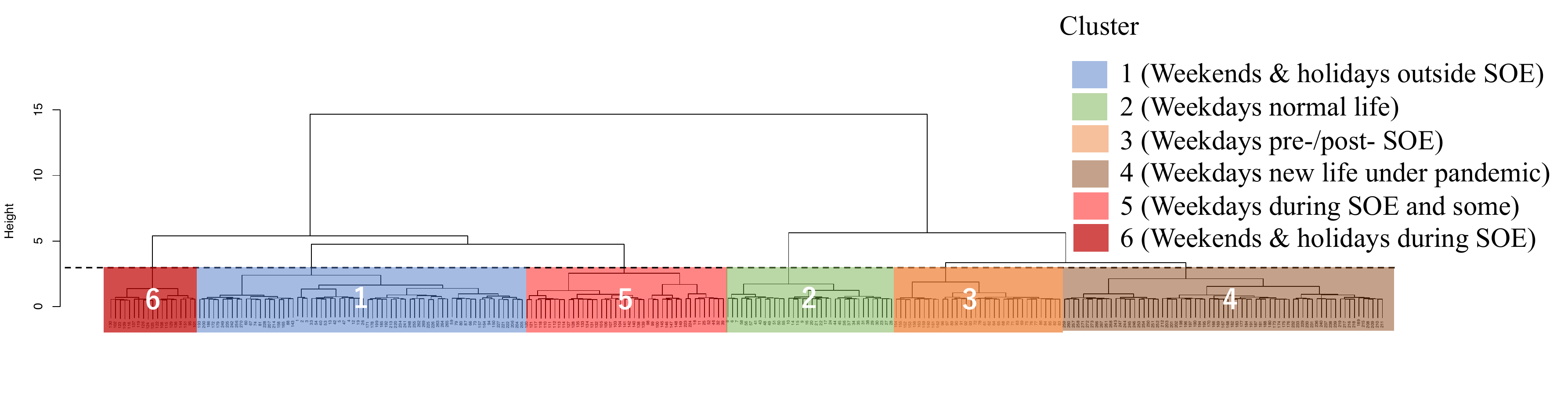}
		\subcaption{O-UEA}
	\end{minipage}
	\caption{Clustering results using dendrograms.}
	\label{fig:hc}
\end{figure}

Looking at the similarities between the clusters in the dendrograms presented in Figure \ref{fig:hc}, the structures of the dendrograms are the same for all three UEAs. The highest branch divides the days into weekends and holidays or weekdays. It can also be seen that the weekends and holidays branch includes weekdays during the SOE. Thus, the mobility patterns for weekdays during the SOE are similar to those for weekends and holidays. This is why cluster 5 (``Weekdays during SOE and some") for the O-UEA includes some weekends outside the SOE. Moreover, the cluster for weekdays during the SOE is more similar to those for normal weekends and weekends with COVID-19 than those for weekends during the SOE for all of the UEAs. 

\subsection{Properties of mobility pattern network}
Next, we investigated the properties of the mobility pattern networks. To do so, we developed representative mobility pattern networks for the clusters by averaging the link weights for the days in each cluster.
\subsubsection{Degree of mobility within residential areas}
 First, because people in Japan were instructed to stay home and work from home, the weights of links to and from residential areas (termed ``inner mobility") were expected to increase during the SOE. It can be seen from Table \ref{tab:percentage_inner_trip} that the clusters for weekends and holidays during the SOE had the highest percentage of inner mobility for all of the UEAs, as expected. The second highest percentage of inner mobility was seen in the cluster of weekends and holidays  pre/post-SOE for the T-UEA, the cluster of weekends and holidays when the situation changed for the N-UEA, and the cluster of weekends/holidays outside the SOE for the O-UEA. As discussed above, people tended to refrain from undertaking outings in cluster 8 (``Weekends \& holidays when situation changed") for the N-UEA. While the cluster of weekdays during the SOE showed the third highest percentage for the T-MEA, it was lower than the cluster of weekends \& holidays under normal life for the N-UEA. This could be because more people worked from home in the T-UEA than in the other UEAs. The survey conducted by the Ministry of Health, Labour and Welfare in conjunction with Line Corporation on 12 and 13 April \citep{MHLW2020} found that while 51.88 \% and 26.28 \% of office workers in Tokyo and Osaka prefectures, respectively, were working from home, only 15.56 \% of office workers in Aichi prefecture were doing so. This low level of teleworking might be one reason for the lower level of inner mobility for the cluster of weekdays during the SOE than for that of holidays under normal conditions. 
We mentioned earlier that people might have felt more able to undertake outings on the days in clusters 9 and 4 for the N-UEA and the O-UEA, respectively. As expected, the level of inner mobility in these clusters was lower than that of the cluster of weekdays just before and after the SOE (i.e., cluster 4 for the N-UEA and cluster 3 for the O-UEA), suggesting that more people spent time in cities other than their city of residence on days distant from the SOE. 

\begin{table}[t]
\centering
\caption{Inner mobility}
\label{tab:percentage_inner_trip}
\begin{tabular}{clclc}
& (a) T-UEA & & (b) N-UEA & \\
Rank & Cluster & \% & Cluster & \%  \\ \hline \hline
1&6 (Weekends\&holidays during SOE)            &86.4   &6 (Weekends\&holidays during SOE)                      &87.4    \\
2&3 (Weekends\&holidays pre/post-SOE)   &83.2   &8 (Weekends\&holidays when situation changed)      &86.1   \\
3&5 (Weekdays during SOE)            &82.5   &3 (Weekends\&holidays pre/post-SOE)      &84.9 \\
4&1 (Weekends\&holidays normal life)          &81.6   &7 (New Year days)                          &84.5 \\
5&4 (Weekdays pre/post-SOE)   &78.2   &1 (Weekends\&holidays normal life)                   &83.7   \\
6&2 (Weekdays normal life)          &75.1   &5 (Weekdays during SOE)                     &83.5  \\
7&                                  &       &4 (Weekdays life pre/post-SOE)      &81.1   \\
8&                                  &       &9 (Weekdays new life under pandemic)      &80.3   \\
9&                                  &       &2 (Weekdays normal life)                   &79.3  \\ 
& (c) O-UEA &\\
Rank & Cluster & \% \\
1   &6 (Weekends\&holidays during SOE)         & 85.8 \\
2   &1 (Weekends\&holidays outside SOE)       & 83.0 \\
3   &5 (Weekdays during SOE and some)& 81.9 \\
4   &3 (Weekdays pre/post-SOE)& 79.2 \\
5   &4 (Weekdays new life under pandemic)& 78.8 \\
6   &2 (Weekdays normal life)       & 77.5\\
\hline
\end{tabular}
\end{table}

\begin{figure}[p]
	\begin{minipage}{0.49\columnwidth}
		\centering
		\includegraphics[width=\columnwidth]{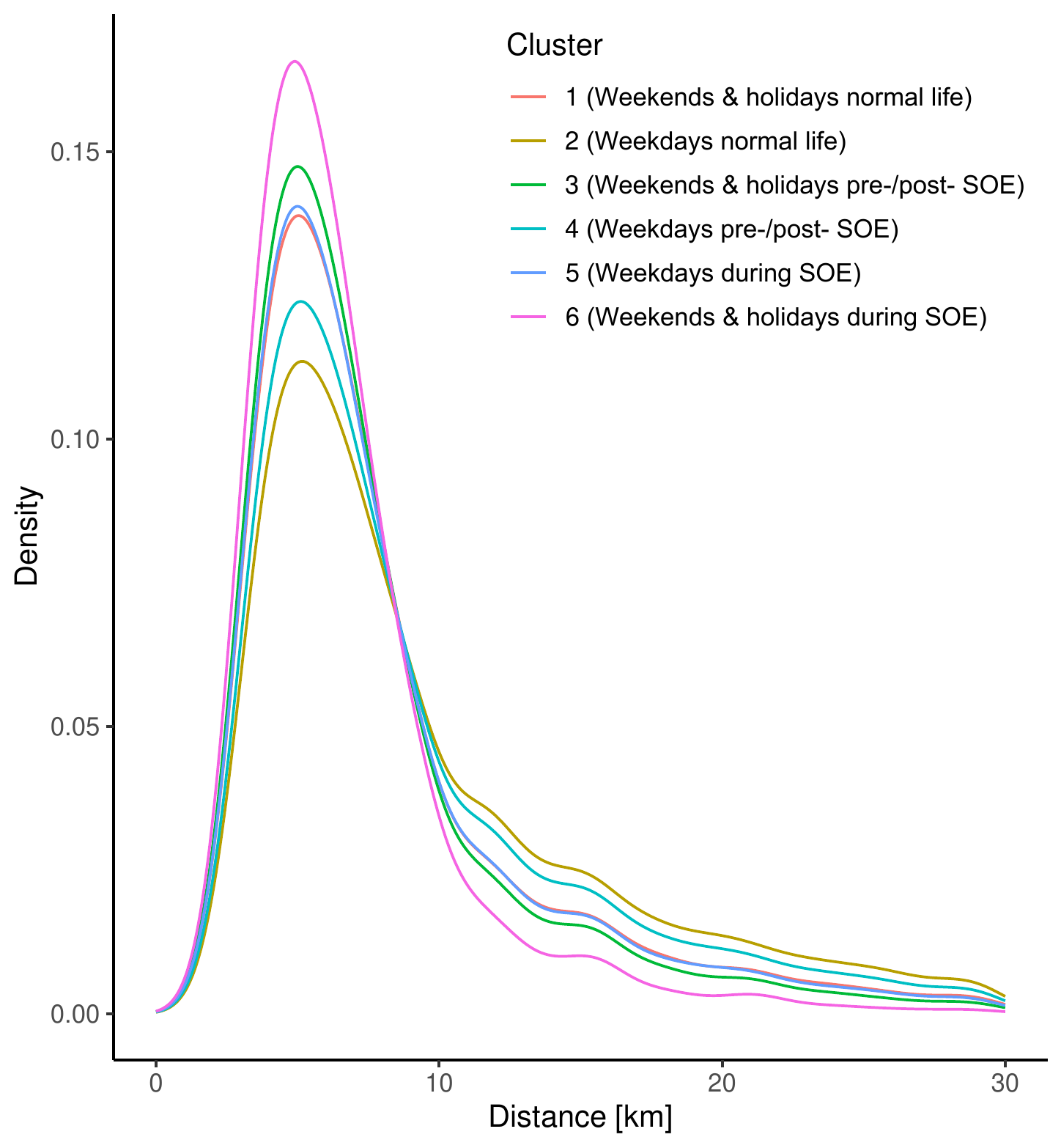}
		\subcaption{T-UEA}
	\end{minipage}
	\begin{minipage}{0.49\columnwidth}
		\centering
		\includegraphics[width=\columnwidth]{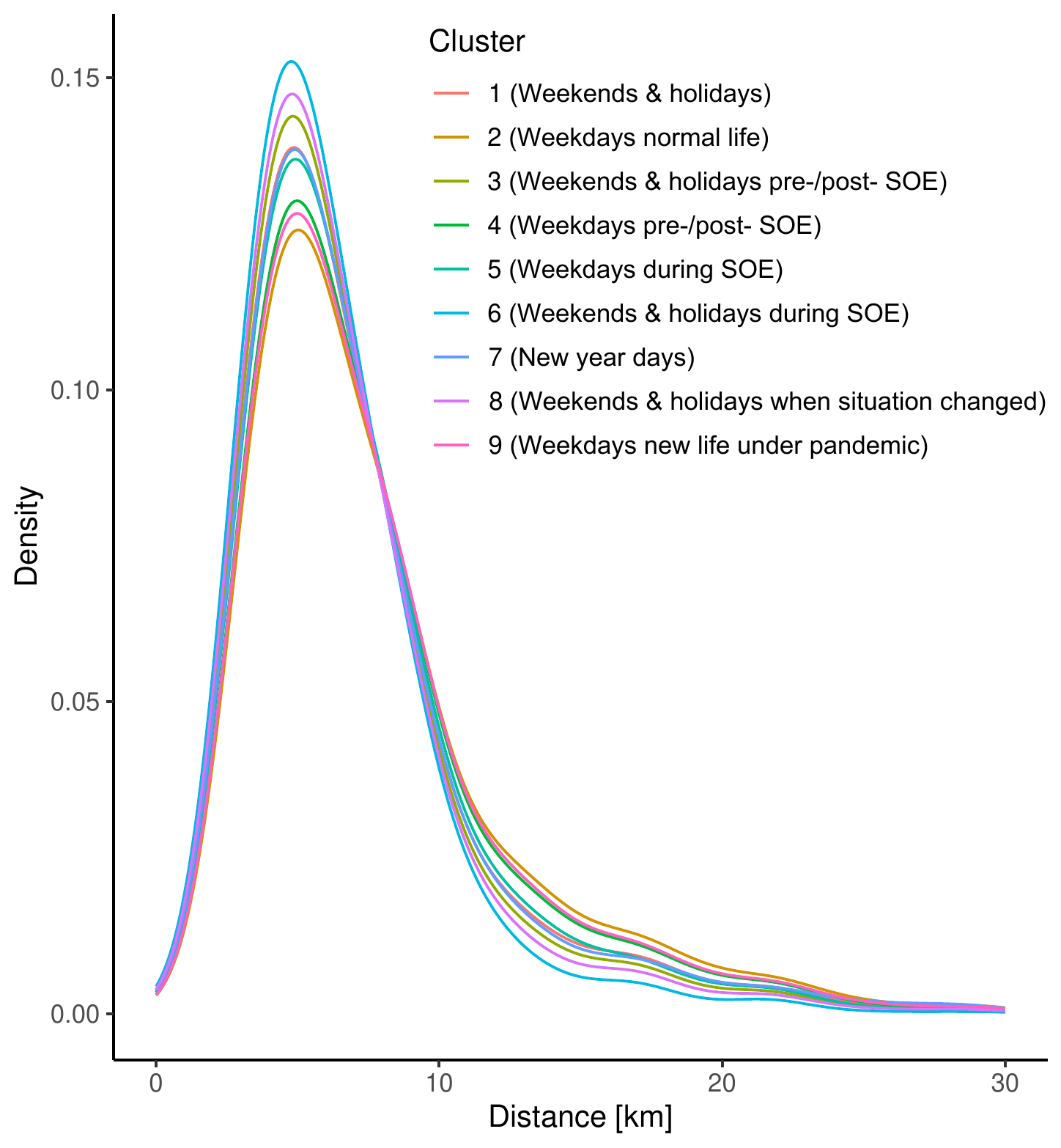}
		\subcaption{N-UEA}
	\end{minipage}
	\begin{minipage}{0.49\columnwidth}
	\centering
		\includegraphics[width=\columnwidth]{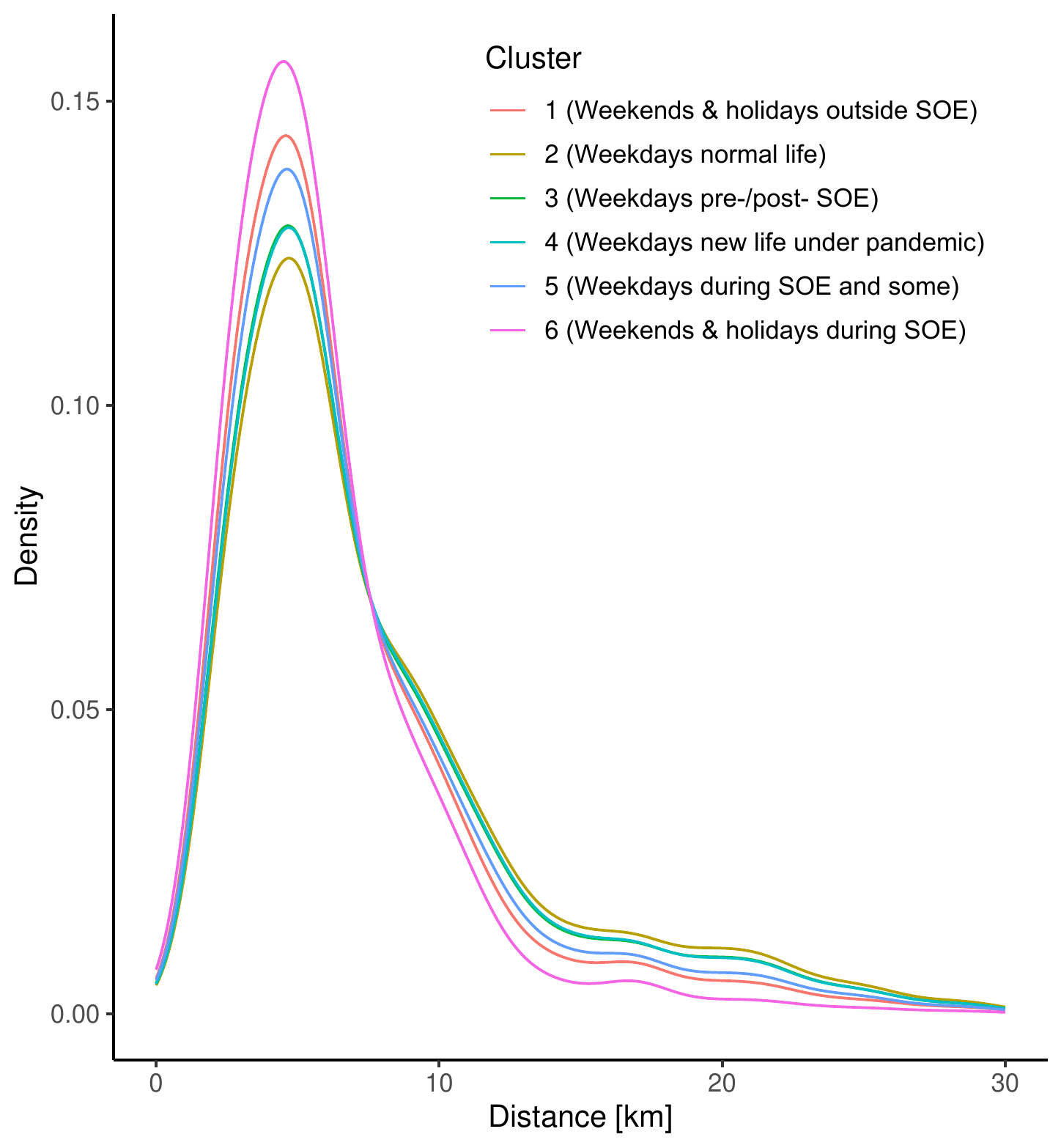}
		\subcaption{O-UEA}
	\end{minipage}
\caption{Trip length distribution ($\leq$30 km). Note that inner mobility has been removed to make the changes in the distribution clear.}
\label{fig:trip_distribution}
\end{figure} 

\subsubsection{Trip length distribution}
To capture how far the cities where people spent their time were located from their city of residence, we estimated the trip length distribution by applying the Kernel density function\footnote{Note that the bandwidth of the Kernel density function is set at 1 [km].}.  Note that the distances between cities are assumed to be the Euclidean distance between the centroids of the cities, and that inner mobility is removed from the data regarding the trip length distribution because the share of inner mobility is quite high, as can be seen from Table \ref{tab:percentage_inner_trip}, which would make the changes in the trip length distribution unclear. The descriptive statistics can be found in Appendix A. 

As shown in Figure \ref{fig:trip_distribution}, there is a peak in the trip length distribution at around $5$ km for all clusters in each of the three UEAs. Remarkably, it can also be seen that the order of density of the distributions is reversed from a distance of around $8$ km for all three UEAs. For instance, cluster 6 (Weekends/holidays during SOE) has the highest peak density for all three UEAs. The densities of these clusters for distances of more than $8$ km were the lowest among the clusters. This would seem to suggest that people refrained from undertaking trips of more than $8$ km. 
 
Regarding weekends and holidays in the T-UEA, the density for short trips (i.e., $\leq 8$ km) of cluster 3 (Weekends \& holidays pre/post-SOE) was higher than that of cluster 1 (Weekends \& holidays normal life), while the density for long trips (i.e., $> 8$ km) was lower. This showed that people started to avoid long trips both before and after the SOE. The same trend can be seen in the N-UEA regarding the distributions of cluster 1 (Weekends \& holidays normal life) and cluster 3 (Weekends \& holidays pre/post-SOE). In addition, the distribution of cluster 8 (Weekends \& holidays when situation changed) is located between that of cluster 1 (Weekends \& holidays normal life) and that of cluster 3 (Weekends \& holidays pre/post-SOE). This is consistent with the abovementioned explanation that people wanted to wait and see what others did when the situation changed. 
 
In terms of weekdays, the distribution of cluster 5 (Weekdays pre/post-SOE) for the T-UEA is similar to that of cluster 1 (Weekends \& holidays normal life), and the share of short trips for cluster 5 is slightly higher than that for cluster 1. The trip length distribution of cluster 4 (Weekdays pre/post-SOE) is between those of cluster 2 (Weekdays normal life) and cluster 5 (Weekdays during SOE), as people had already started to refrain from undertaking outings before the SOE and continued to do so even after the SOE was lifted, although they were not as strict as during the SOE. The same situation occurred in relation to the other two UEAs. It can be seen in relation to the N-UEA that because the density of cluster 5 (Weekdays during SOE) for short trips is lower than those of cluster 1 (Weekends \& holidays normal life) and cluster 3 (Weekends \& holidays pre/post-SOE), people undertook more outings on weekdays during the SOE than on normal weekends and before and after the SOE. This could also be because of the lower level of teleworking that occurred in the N-UEA, that is, people had to undertake longer trips to get to work. The trip length distribution of cluster 9 (Weekdays new life under pandemic) is between those of cluster 2 (Weekdays normal life) and cluster 4 (Weekdays pre/post-SOE). Thus, it can be said that after the SOE was lifted, people gradually started to venture out, and life returned to a similar situation to that which existed prior to the COVID-19 outbreak. Conversely, in the case of the O-UEA, the trip distributions for cluster 3 (Weekdays pre/post-SOE) and cluster 4 (Weekdays new life under pandemic) were similar, with the only difference possibly being related to the different levels of inner mobility noted earlier. 

  \begin{figure}[h]
	\begin{minipage}{0.5\columnwidth}
		\centering
		\includegraphics[width=\columnwidth]{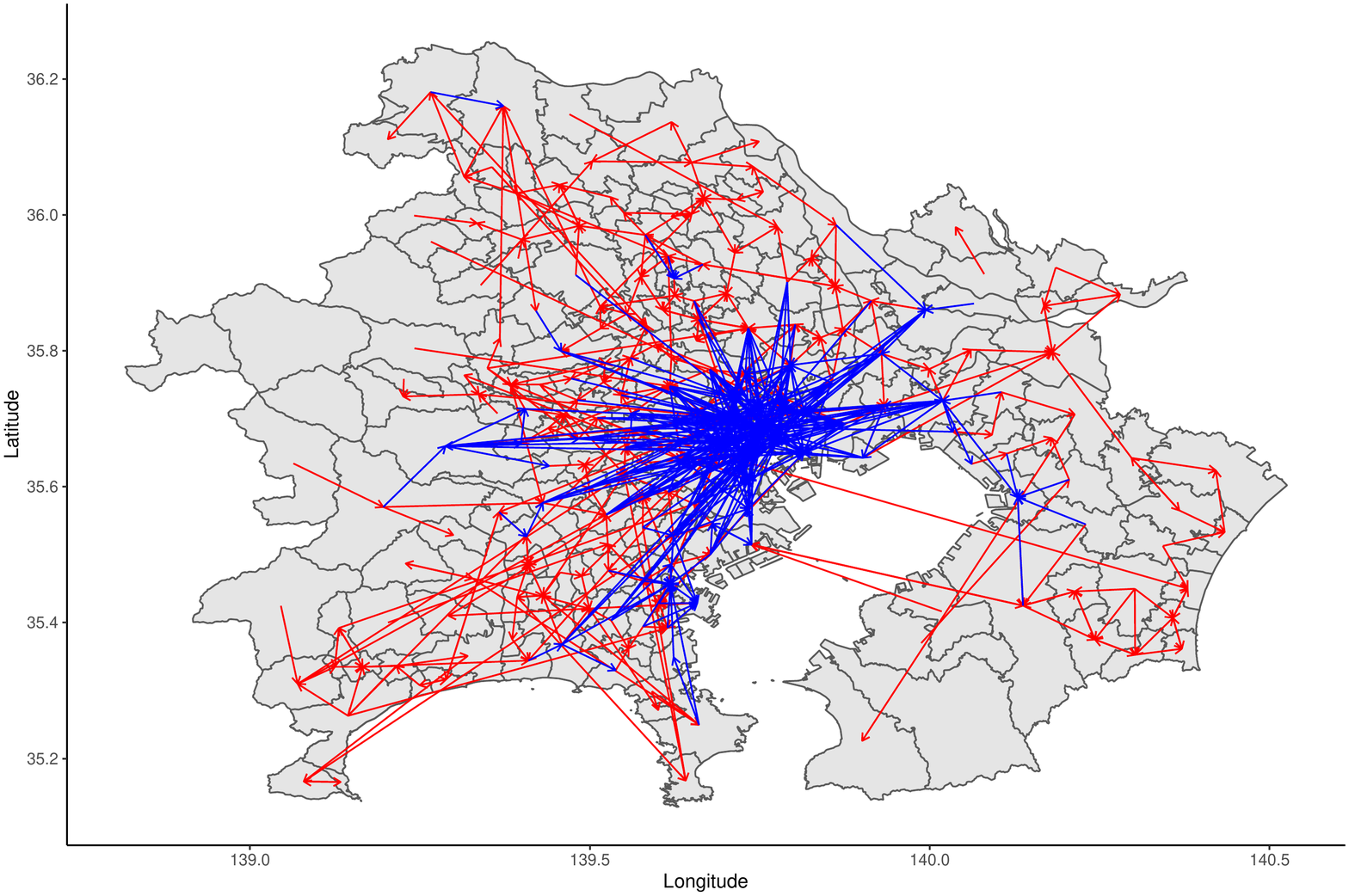}
		\subcaption{T-UEA}
		\end{minipage}
	\begin{minipage}{0.5\columnwidth}
	\centering
		\includegraphics[width=\columnwidth]{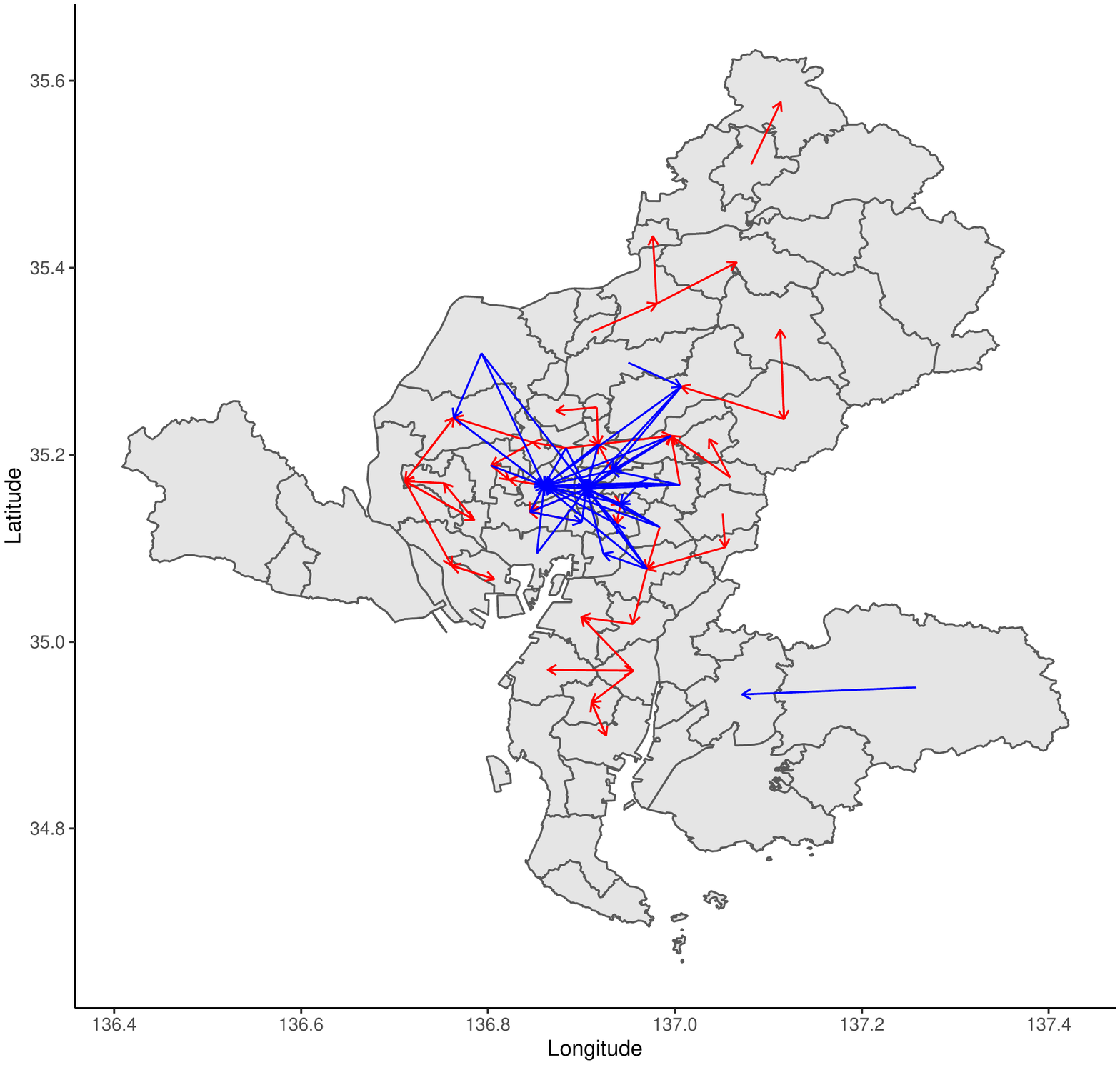}
		\subcaption{N-UEA}
	\end{minipage}
		\begin{minipage}{0.5\columnwidth}
	\centering
		\includegraphics[width=\columnwidth]{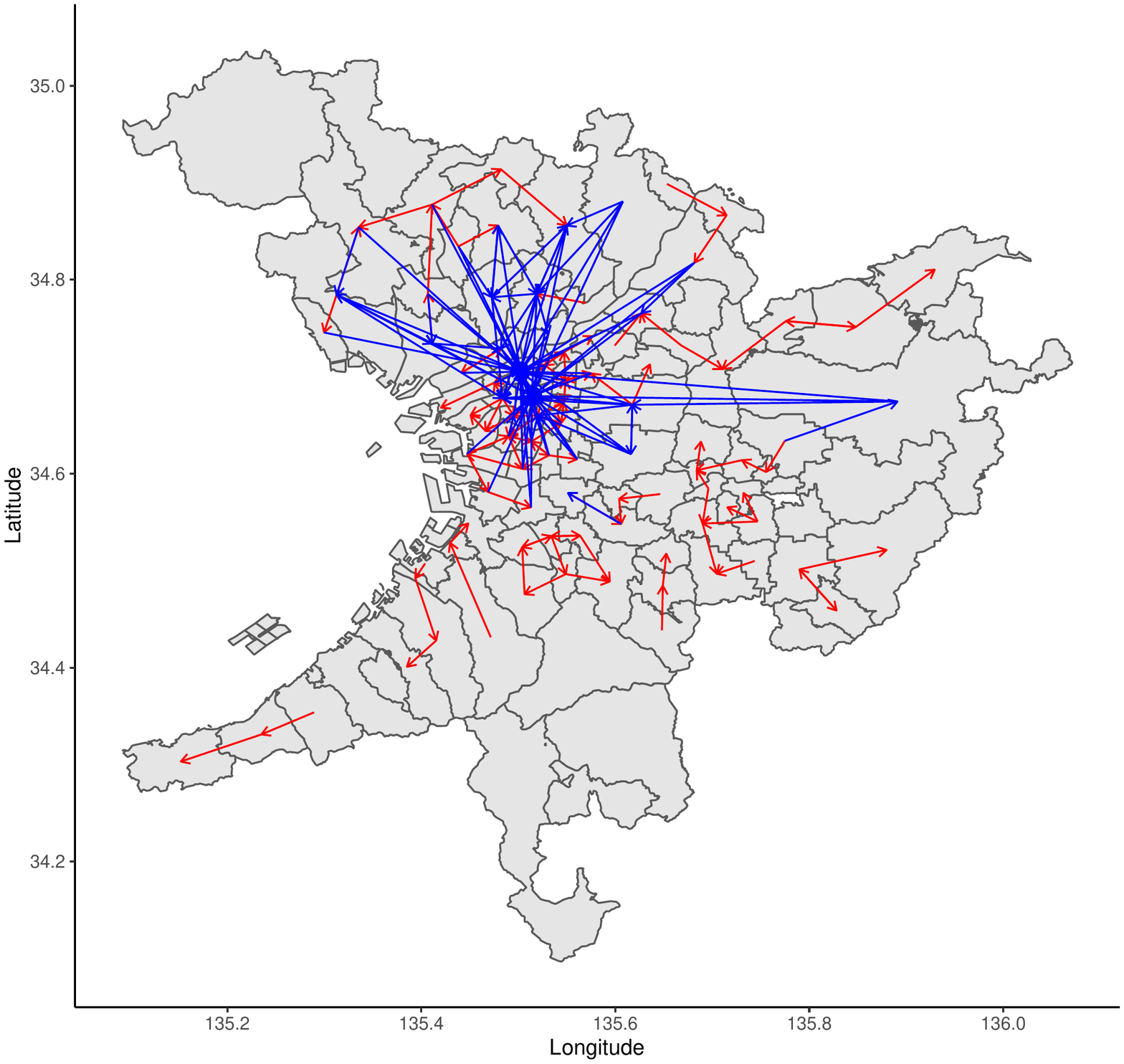}
		\subcaption{O-UEA}
	\end{minipage}
\caption{Spatial distribution of the highest changes in the weights of the links from the representative mobility pattern network for normal weekdays to that for weekdays during the SOE (blue: 99 percentile reduction in weight, red: 99 percentile increase in weight).}
\label{fig:spatial_distribution}
\end{figure} 

\subsubsection{Spatial properties of mobility pattern networks}
We showed earlier that inner mobility increased during the SOE in all three UEAs because people stayed home and worked from home. Next, we investigate how the mobility pattern networks for weekdays changed spatially, as shown in Figure \ref{fig:spatial_distribution}. We use clusters 2 and 5 for each of the three UEAs. The blue and red lines represent links that experienced a 99 percentile reduction in weight and a 99 percentile increase in weight, respectively, from the network for normal weekdays during normal life to that during the SOE. A reduction in mobility can be seen for links to downtown cities such as Chiyoda ward for the T-UEA, Nagoya city for the N-UEA, and Osaka city for the O-UEA because people worked from home and avoided going to these areas, which are normally heavily populated. Conversely, mobility to adjacent cities increased. This could be because instead of going to downtown areas, people went to places such as parks and shopping areas near their homes. 

\section{Conclusion}
We investigated changes in travel patterns using longitudinal mobile phone data to construct mobility pattern networks. The results for the three largest urban employment areas (i.e., Tokyo, Nagoya, and Osaka) revealed some interesting findings. First, by applying hierarchical clustering to the distance matrices, we identified change points in the mobility pattern networks. We found that the mobility pattern networks changed at the beginning of March when schools were ordered to close. We also found that the mobility pattern networks during the SOE differed significantly from those in other periods.  Additionally, our results suggest that the mobility patterns did not change for several days after the SOE was lifted.  Even though several policy measures were implemented by some local governments, such as an alert by the Tokyo local government and the self-declared SOE by the Aichi local government, changes in mobility patterns had already occurred prior to these measures. Thus, it seems that the Japanese people voluntarily changed their behavior not only in response to policy measures but also on the basis of information regarding the number of COVID-19 cases. Second, we analyzed how the mobility patterns changed between clusters. It can be seen that because more people stayed home and worked from home during the SOE, inner mobility increased and the weights of the links from cities of residence to other cities were lower than those of the other clusters. However, more people traveled beyond their city of residence on weekdays during the SOE than on weekends before and after the SOE in the N-UEA and the O-UEA, which could be because of the lower levels of teleworking in the N-UEA and the O-UEA. Thus, the ability to telework could be an important factor in controlling the movement of people, as indicated in \cite{di2020expected}. Third, the spatial properties of the mobility pattern networks revealed that most of the reduction in mobility from weekdays during normal life to those during the SOE could be seen at the links to the downtown areas, while mobility increased in relation to links to adjacent cities. This finding is consistent with the fact that people refrained from traveling, especially to busy downtown areas where people normally like to go. 

As mentioned above, our analyses revealed some mobility pattern changes that cannot be captured using raw mobile phone data. Thus it is useful to investigate how and when mobility patterns changed in response to the COVID-19 outbreak and the related policy measures. Several directions remain for future work. First, the situation regarding the COVID-19 pandemic remains unpredictable. Thus, it is necessary to continue to monitor changes in mobility patterns. Second, our analysis only focused on three areas, and thus it would be interesting to expand the scope of the study to include the entire country. 

\section*{Declaration of Competing Interest}
The authors declare that they have no known competing financial interests or personal relationships that could have appeared to influence the work reported in this paper.

\section*{Acknowledgments}
We would like to thank DOCOMO Insight Marketing, Inc. for providing aggregated mobile phone data termed Mobile Spatial Statistics. This study was supported by a Grant-in-Aid for Scientific Research from the Japan Society for the Promotion of Science (B) \# 20H02269 and a Grant-in-Aid for Research Activity start-up \# 19K23531. We would also like to thank Satoru Kimura at the Tokyo Institute of Technology for helping with the data processing.

\section*{Appendix A}
\setcounter{table}{0}
\renewcommand{\thetable}{A\arabic{table}}
In this section, we provide a descriptive statistics of trip lengths discussed in section 4.4.2. It is important again to note that the inner mobility is removed from the data. Mean and percentiles of each cluster is summarized in Table.\ref{tab:descriptive}.

\begin{table}[b]
\centering
\caption{Descriptive Statistics of trip lengths}
\label{tab:descriptive}
\begin{tabular}{llcccccc}
&&& \multicolumn{5}{c}{Percentile} \\ \cline{4-8}
UEA & Cluster & Mean & 0 &  25 & 50 & 75 & 100 \\
\hline
T-UEA & 1 (Weekends \& holidays normal life) & 9.13 & 2.60 & 4.67 & 6.56 & 9.89 & 132.05 \\
& 2 (Weekdays normal life) & 10.92& 2.60 & 5.21 & 7.71 & 13.49 & 129.22  \\
& 3 (Weekends \& holidays pre-/post- SOE) & 8.26& 2.60 & 4.58 & 6.21 & 9.03 & 135.08 \\
& 4 (Weekdays pre-/post- SOE) & 9.93& 2.60 & 4.93 & 7.13 & 11.86 & 132.05 \\
& 5 (Weekdays during SOE) & 8.65 & 2.60 & 4.63 & 6.39 & 9.55 & 127.09 \\ 
& 6 (Weekends \& holidays during SOE) & 6.98 & 2.60 & 4.37 & 5.78 & 7.84 & 131.24 \\
N-UEA & 1 (Weekends \& holidays) & 7.50 & 2.40 & 4.24 & 5.92 & 8.46 & 95.89 \\ 
& 2 (Weekdays normal life) & 8.19 & 2.40 & 4.62 & 6.43 & 9.34 & 84.44 \\ 
& 3 (Weekends \& holidays pre-/post- SOE) & 7.16 & 2.40 & 4.19 & 5.88 & 8.17 & 88.09 \\
& 4 (Weekdays pre-/post- SOE) & 7.80 & 2.40 & 4.40 & 6.28 & 8.96 & 88.09 \\ 
& 5 (Weekdays during SOE) & 7.30 & 2.40 & 4.30 & 5.92 & 8.46 & 84.64 \\ 
& 6 (Weekends \& holidays during SOE) & 6.44 & 2.40 & 4.06 & 5.36 & 7.67 & 84.64 \\ 
& 7 (New Year holidays) & 7.65 & 2.40 & 4.30 & 5.92 & 8.47 & 95.89 \\ 
& 8 (Weekends \& holidays when situation changed) & 6.90 & 2.40 & 4.14 & 5.72 & 8.07 & 84.64 \\ 
& 9 (Weekdays new life under pandemic) & 7.91 & 2.40 & 4.45 & 6.36 & 9.08 & 91.02 \\
O-UEA & 1 (Weekends \& holidays outside SOE) & 6.95 & 1.92 & 3.88 & 5.35 & 10.11 & 89.22 \\ 
& 2 (Weekdays normal life) & 8.34 & 1.92 & 4.26 & 5.97 & 10.11 & 80.49 \\ 
& 3 (Weekdays pre-/post- SOE) & 7.91 & 1.92 & 4.20 & 5.80 & 9.55 & 83.86 \\
& 4 (Weekdays new life under pandemic) & 7.93 & 1.92 & 4.25 & 5.81 & 9.55 & 83.86 \\ 
& 5 (Weekdays during SOE and some) & 7.26 & 1.92 & 3.97 & 5.55 & 8.64 & 83.86 \\ 
& 6 (Weekends \& holidays during SOE) & 6.02 & 1.92 & 3.58 & 5.07 & 7.08 & 83.86 \\
\hline
&&&&&&& Unit: km
\end{tabular}
\end{table}

\bibliographystyle{elsarticle-harv}
\bibliography{ref}
\end{document}